\documentclass[prd,amsmath,amssymb,reprint,preprintnumbers,nofootinbib,superscriptaddress]{revtex4-1}
              
\pdfoutput=1

\usepackage{amsmath}
\usepackage{amsfonts}
\usepackage{amssymb}
\usepackage{mathrsfs}
\usepackage{graphicx}
\usepackage{color}
\usepackage[dvipsnames]{xcolor}
\usepackage{longtable}
\usepackage{bm}
\usepackage{blindtext}
\usepackage{wasysym}
\usepackage{hyperref}
\hypersetup{colorlinks=true,allcolors=blue}
\usepackage[normalem]{ulem}
\usepackage{lineno}

\newcommand\Fontx{\fontsize{10}{12}\selectfont}

\begin{document}
 
\title{Exploring constraints on the core radius and density jumps inside Earth using atmospheric neutrino oscillations\\\vskip0.2cm
\Fontx{``Contribution to the 25th International Workshop on Neutrinos from Accelerators''}}
\author{Anuj Kumar Upadhyay}
\email{anuju@iopb.res.in (ORCID: 0000-0003-1957-2626)}
\affiliation{Department of Physics, Aligarh Muslim University, Aligarh 202002, India}
\affiliation{Institute of Physics, Sachivalaya Marg, Sainik School Post, Bhubaneswar 751005, India}
\author{Anil Kumar}
\affiliation{Institute of Physics, Sachivalaya Marg, Sainik School Post, Bhubaneswar 751005, India}
\affiliation{Applied Nuclear Physics Division, Saha Institute of Nuclear Physics, Bidhannagar, Kolkata 700064, India}
\affiliation{Homi Bhabha National Institute, Anushakti Nagar, Mumbai 400094, India}
\author{Sanjib Kumar Agarwalla}
\affiliation{Institute of Physics, Sachivalaya Marg, Sainik School Post, Bhubaneswar 751005, India}
\affiliation{Homi Bhabha National Institute, Anushakti Nagar, Mumbai 400094, India}
\affiliation{Department of Physics \& Wisconsin IceCube Particle Astrophysics Center, University of Wisconsin, Madison, WI 53706, U.S.A}
\author{Amol Dighe} 
\affiliation{Tata Institute of Fundamental Research, Homi Bhabha Road, Colaba, Mumbai 400005, India}

\preprint{NuFact 2024-17}

\date{\today}

\begin{abstract}
	
\begin{center} (presented by Anuj Kumar Upadhyay)\end{center}

Atmospheric neutrinos, through their weak interactions, can serve as an independent tool for exploring the internal structure of Earth. The information obtained would be complementary to that provided by seismic and gravitational measurements. The Earth matter effects in neutrino oscillations depend upon the energy of neutrinos and the electron density distribution that they encounter during their journey through Earth, and hence, can be used to probe the inner structure of Earth. In this contribution, we demonstrate how well an atmospheric neutrino experiment, such as an iron calorimeter detector (ICAL), would simultaneously constrain the density jumps inside Earth and determine the location of the core-mantle boundary. In this work, we employ a five-layered density model of Earth, where the layer densities and core radius are modified to explore the parameter space, ensuring that the mass and moment of inertia of Earth remain constant while satisfying the hydrostatic equilibrium condition. We further demonstrate that the charge identification capability of an ICAL-like detector would play a crucial role in obtaining these correlated constraints.
  
\end{abstract}


\maketitle


\section{Introduction} 

The knowledge about the interior of Earth is primarily derived from the indirect probes such as seismic studies and gravitational measurements. The widely used density model of Earth, known as the Preliminary Reference Earth Model (PREM)~\cite{Dziewonski:1981xy}, has been developed using the seismic wave propagation data along with the gravitational constraints on Earth's total mass and moment of inertia. These indirect probes have significantly improved our understanding of the inner structure of Earth. Still, we have some open questions. Complementary and independent information about the internal structure of Earth can be obtained by observing the upward-going atmospheric neutrinos that travel through Earth. As multi-GeV atmospheric neutrinos traverse through Earth, they experience the Earth's matter effects~\cite{Wolfenstein:1977ue,Mikheev:1986gs} due to the coherent and forward charged-current interactions with the ambient electrons. These matter effects modify the oscillation probabilities of neutrinos travelling through Earth. The matter effects depend upon the sudden density jump at the core-mantle boundary (CMB) and its location. Therefore, any modification in the density jump, CMB location, or both simultaneously, will lead to changes in the neutrino oscillation patterns.

In this work, we study the effect of modifications in the density jump at CMB and its location on the neutrino oscillation patterns using the simulated data for the proposed 50~kt Iron Calorimeter (ICAL) detector at the India-based Neutrino Observatory (INO)~\cite{ICAL:2015stm}. We further demonstrate that the capability of the ICAL detector to distinguish neutrinos from antineutrinos would play an important role in effectively constraining the parameter space. This contribution is based on Ref.~\cite{Upadhyay:2024gra}.

\section{Density Profiles of Earth} 

\begin{figure*}
	\centering
	\includegraphics[width=0.84\linewidth]{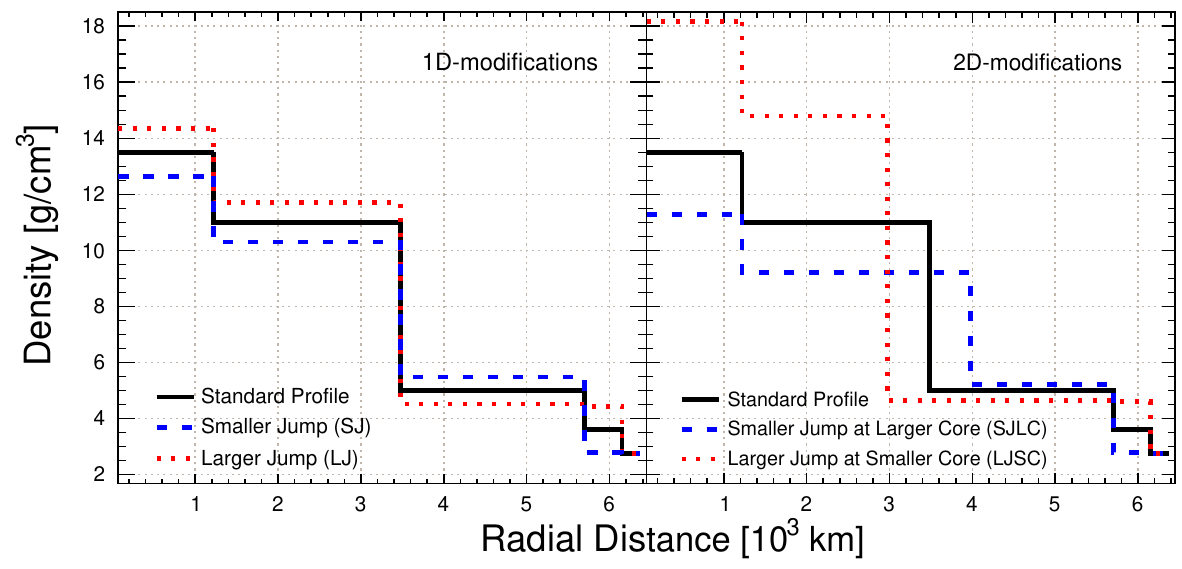}
	\caption{Alternative five-layered density profiles of Earth as a function of the radial distance. The black curves denote the standard density profile with a density jump of $\Delta \rho_\text{CMB} = 6.0$ g/cm$^3$ at the CMB, located at its standard radius of $R_\text{CMB} = 3480$~km. The left panel illustrates the 1D modifications, where dashed-blue (dotted-red) curve corresponds to the SJ (LJ) scenario. The right panel shows the 2D modifications, where the dashed-blue (dotted-red) curve represents the SJLC (LJSC) scenario. See text for the definition of SJ, LJ, SJLC, and LJSC. This figure is taken from Ref.~\cite{Upadhyay:2024gra}.} 
	\label{fig:dend_profile}
\end{figure*} 

In this study, we use a five-layered density model of Earth, where the density within each layer is uniform. This density model is a simplified form of the PREM~\cite{Dziewonski:1981xy} profile. The five distinct layers are the inner core (IC), the outer core (OC), the inner mantle (IM), the middle mantle (MM), and the outer mantle (OM), with their respective densities represented by $\rho_\text{IC}$, $\rho_\text{OC}$, $\rho_\text{IM}$, $\rho_\text{MM}$, and $\rho_\text{OM}$, respectively. In this model, the following four significant density jumps occur at different layer boundaries: (i) density jump at the inner core - outer core boundary, represented by $\Delta\rho_\text{IC-OC} = \rho_\text{IC} - \rho_\text{OC}$, (ii) density jump at the outer core - inner mantle boundary, also known as CMB, represented by $\Delta\rho_\text{CMB} = \rho_\text{OC} - \rho_\text{IM}$, (iii) density jump at the inner mantle - middle mantle boundary, represented by $\Delta\rho_\text{IM-MM} = \rho_\text{IM} - \rho_\text{MM}$, and (iv) density jump at the middle mantle - outer mantle boundary, represented by $\Delta\rho_\text{MM-OM} = \rho_\text{MM} - \rho_\text{OM}$. The black curve in Fig.~\ref{fig:dend_profile} represents this five-layered density profile of Earth.

We modify the layer densities and the CMB radius to explore the parameter space while ensuring the following constraints: (i) the total mass and moment of inertia of Earth are fixed, (ii) the density of OM ($\rho_\text{OM}$) is fixed, (iii) the radii of the IC, IM, MM, and Earth are taken to be fixed at their standard values, and (iv) the density ratio of the IC and OC ($\rho_\text{IC}/\rho_\text{OC}$) is taken to be the same as their ratio in the PREM profile. We also impose an additional constraint of the hydrostatic equilibrium condition where the density of any inner layer is always greater than that of the outer layer ($\rho_\text{inner layer} > \rho_\text{outer layer}$). With these eight constraints, we are left with two free parameters out of ten parameters. We choose these two free parameters to be $R_\text{CMB}$ and $\Delta\rho_\text{CMB}$. One of the other density jumps, such as $\Delta\rho_\text{IC-OC}$, $\Delta\rho_\text{IM-MM}$, and $\Delta\rho_\text{MM-OM}$, can also be chosen as the second free parameter. In the results section, we will also interpret sensitivities in terms of these other density jumps.

The five-layered model with the constraints mentioned above allows us to modify these two free parameters in the following two ways: (i) 1-dimensional (1D) modification, where only the $\Delta\rho_\text{CMB}$ is varied at fixed $R_\text{CMB} = 3480$~km and (ii) 2-dimensional (2D) modification, where both $\Delta\rho_\text{CMB}$ and $R_\text{CMB}$ are varied simultaneously. The left panel of Fig.~\ref{fig:dend_profile} shows the density profile for some representative choice of densities of the 1D-modifications, where the dashed-blue (dotted-red) curve corresponds to the smaller jump or SJ (larger jump or LJ). The right panel of Fig.~\ref{fig:dend_profile} presents the density profile for some representative choice of densities of the 2D-modifications, where the dashed-blue (dotted-red) curve corresponds to the smaller jump at larger core or SJLC (larger jump at smaller core or LJSC). The details of these scenarios can be found in Ref.~\cite{Upadhyay:2024gra}.

\section{Results}

In this section, we present the expected median sensitivity of the ICAL detector for determining the correlated constraints on the density jumps and the location of CMB. In order to quantify it, we simulate the prospective data obtained using the INO-ICAL detector assuming the standard five-layered density profile of Earth. The unoscillated neutrino events are generated using the NUANCE Monte Carlo (MC) neutrino event generator with the Honda atmospheric neutrino flux (3D) at the INO site. All the details related to the simulation are outlined in Ref.~\cite{ICAL:2015stm}. For this study, we utilize the 20-year of MC data for the 50~kt ICAL detector which is equivalent to 1~Mt$\cdot$yr exposure. We binned the MC data using a binning scheme as given in Table 4 in Ref.~\cite{Upadhyay:2022jfd}. The expected median sensitivity is calculated in terms of $\chi^2$ by fitting the prospective data with the modified density jump at CMB and its location in theory. We follow the definition of $\Delta\chi^2_\text{1D-DJ}$ for the 1D modifications and $\Delta\chi^2_\text{DJ-CMB}$ for the 2D modifications as described in Ref.~\cite{Upadhyay:2024gra}.

\begin{figure*}
	\centering
	\includegraphics[width=0.48\linewidth]{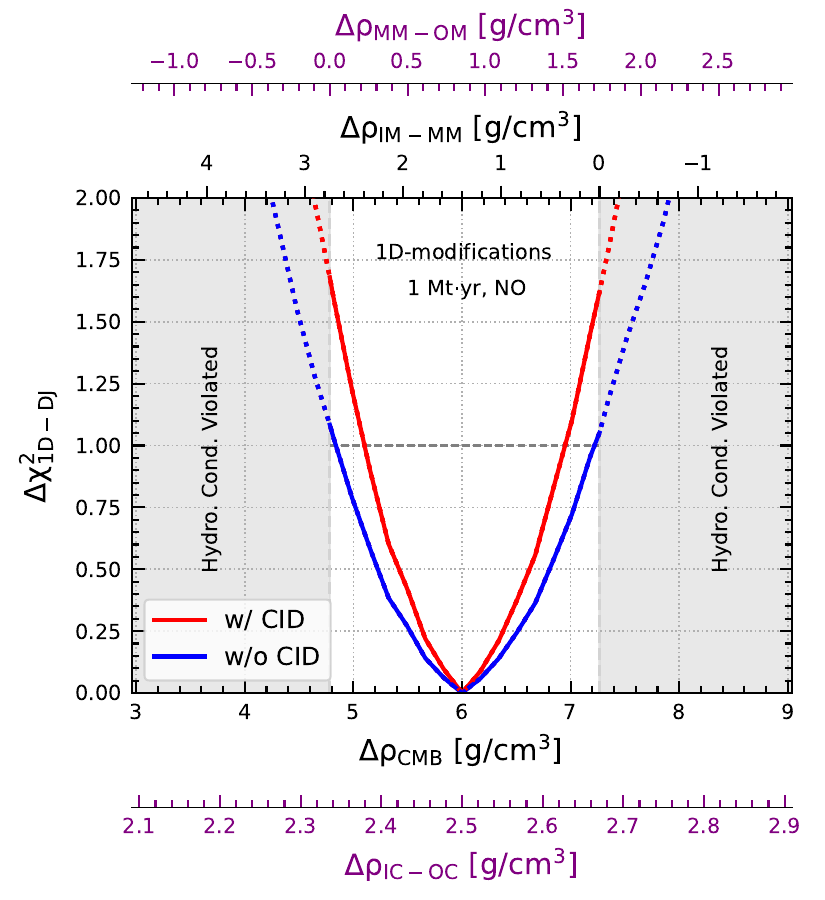}
	\includegraphics[width=0.48\linewidth]{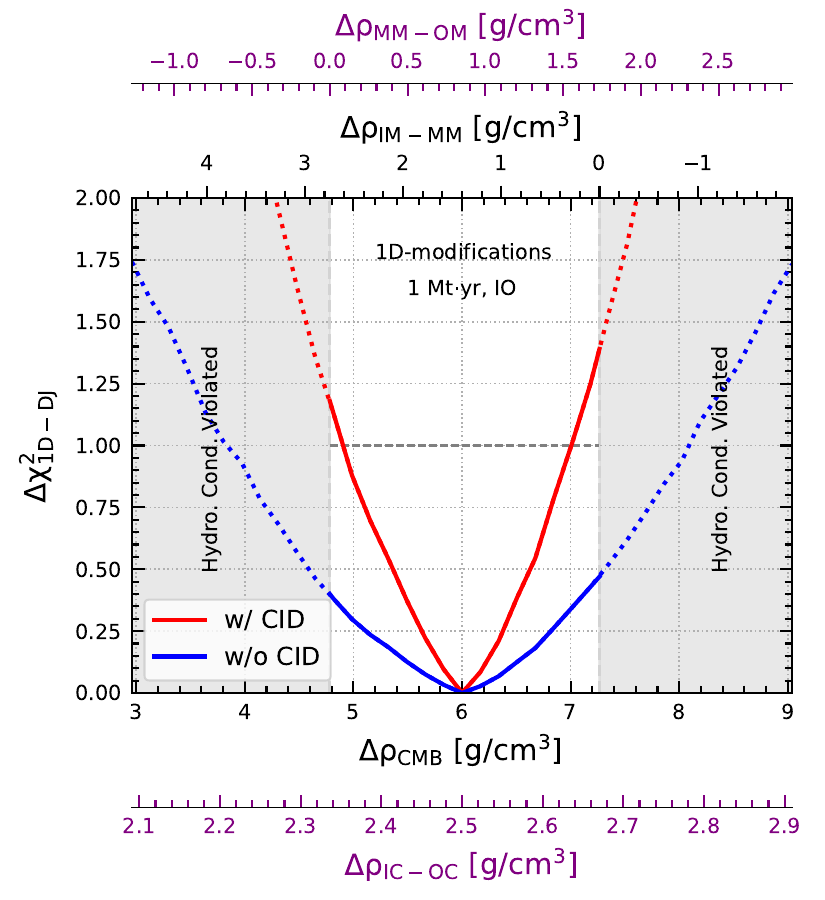}
	\caption{The expected median $\Delta\chi^2_\text{1D-DJ}$ as a function of the density jumps at the respective boundaries, assuming the standard CMB location (1D modifications), for 1~Mt$\cdot$yr exposure of ICAL. The top (bottom) two x-axes correspond to the density jumps at the MM-OM and IM-MM (CMB and IC-OC) boundaries. The gray regions highlight the unphysical area where the hydrostatic equilibrium condition is violated. The left (right) panel illustrates the sensitivity assuming the true neutrino mass ordering to be normal (inverted). The red (blue) curve presents the sensitivity with (without) the CID capability of the ICAL detector. This figure is taken from Ref.~\cite{Upadhyay:2024gra}.}
	\label{fig:1D_results}
\end{figure*}

Figure~\ref{fig:1D_results} presents the expected median sensitivity of the ICAL detector in terms of $\Delta\chi^2_\text{1D-DJ}$ as a function of correlated density jumps at the IC-OC, OC-IM (CMB), IM-MM, and MM-OM boundaries with an exposure of 1~Mt$\cdot$yr. The top (bottom) two x-axes represent the density jumps at the MM-OM and IM-MM (CMB and IC-OC) boundaries. The gray areas highlight the unphysical regions where the hydrostatic equilibrium condition is violated. The left and right panels of Fig.~\ref{fig:1D_results} correspond to the expected sensitivity, assuming the true neutrino mass ordering to be normal and inverted (NO and IO), respectively. In both the panels, the red (blue) curve shows the expected sensitivity with (without) the charge identification or CID capability of the ICAL detector. If the true neutrino mass ordering is NO (IO), the ICAL detector would be able to constrain the value of the density jump at the standard CMB location $\Delta\rho_\text{CMB}$ to be [5.1, 7.0] ([4.9, 7.0]) g/cm$^3$ at $1 \sigma$ in the presence of CID capability which correspond to a precision of about 16\% (18\%). As mentioned, the CID capability of the ICAL detector would play a crucial role in achieving this precision. Without the CID capability, the $1\sigma $ precision would deteriorate to about 20\% for NO. In the absence of the CID capability, the sensitivity gets reduced significantly if the true neutrino mass ordering is IO.

\begin{figure*}
	\centering
	\includegraphics[width=0.45\linewidth]{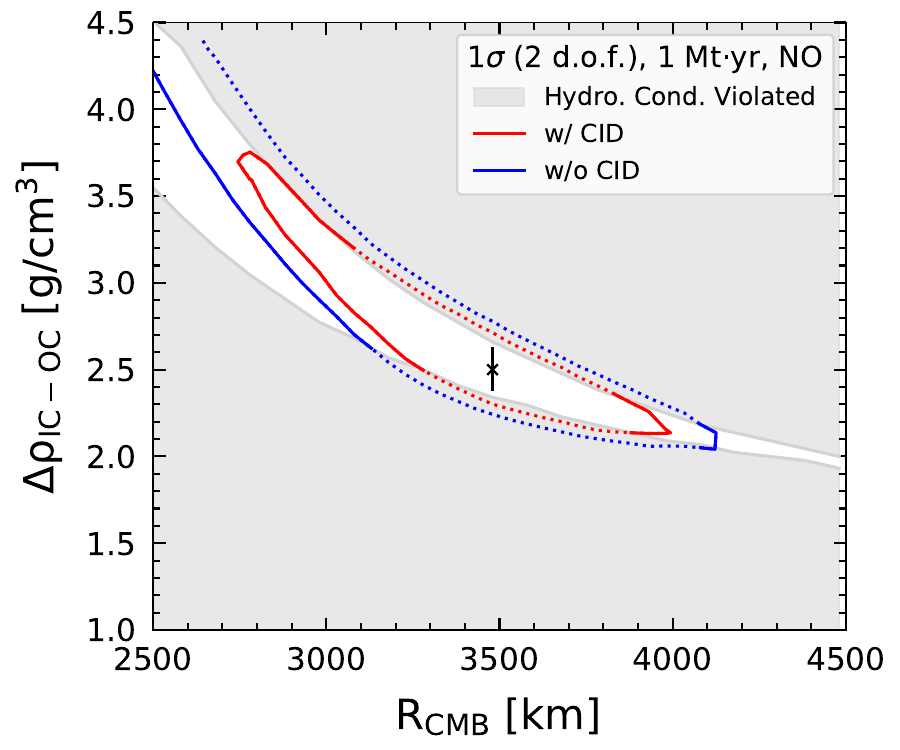}
	\includegraphics[width=0.45\linewidth]{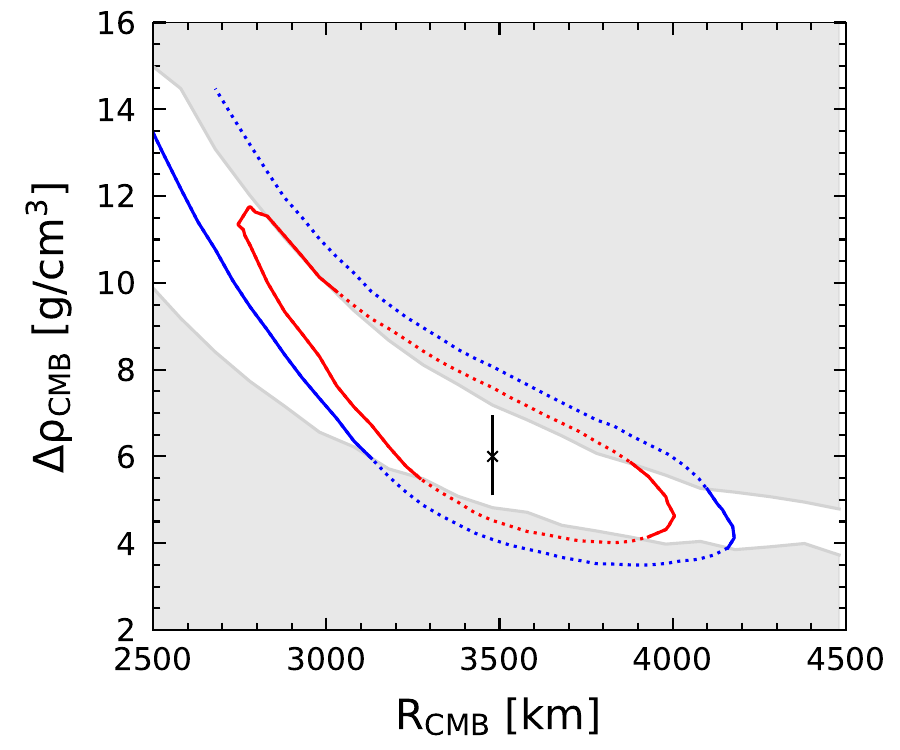}
	\includegraphics[width=0.45\linewidth]{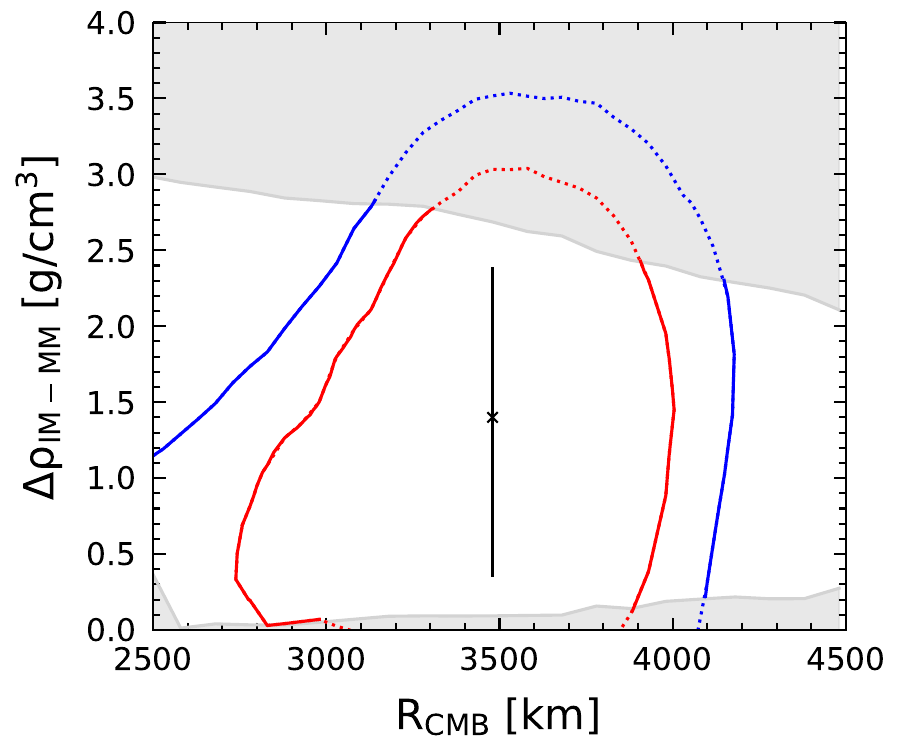}
	\includegraphics[width=0.45\linewidth]{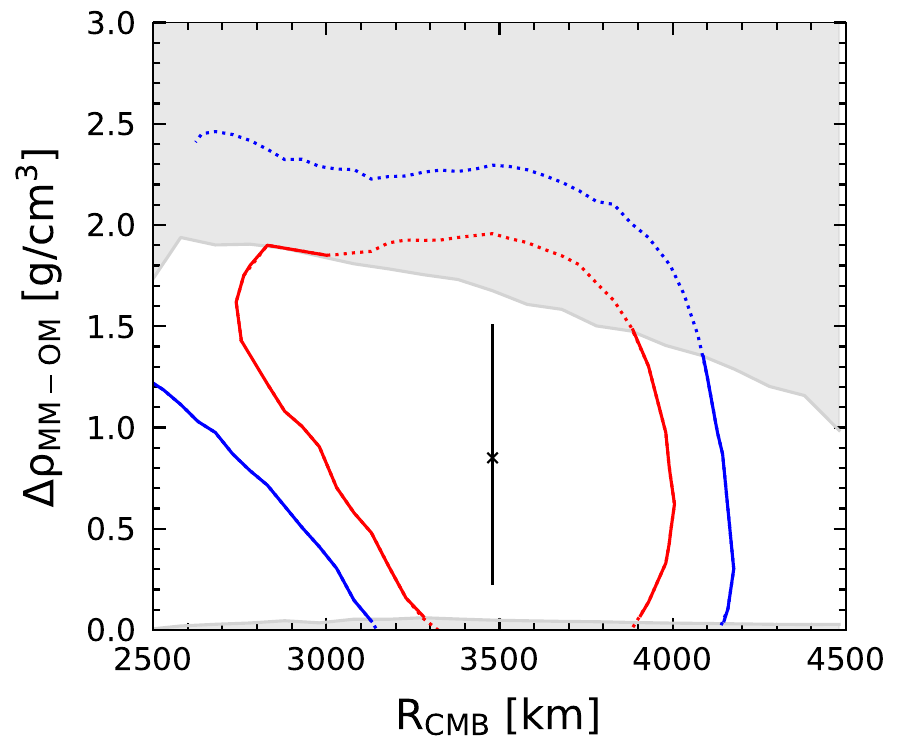}
	\caption{The expected median $\Delta\chi^2_\text{DJ-CMB}$ sensitivity contours in the plane of density jumps and $R_\text{CMB}$ at $1\sigma$ (2 d.o.f.) for 1~Mt$\cdot$yr exposure of ICAL. The red (blue) contour shows the sensitivity with (without) the CID capability of the ICAL detector. The gray regions highlight the unphysical part where the hydrostatic equilibrium condition is violated. These contours correspond to the sensitivity assuming the true neutrino mass ordering to be normal. This figure is taken from Ref.~\cite{Upadhyay:2024gra}.}
	\label{fig:2D_results}
\end{figure*}

Figure~\ref{fig:2D_results} shows the expected sensitivity of the ICAL detector for measuring the correlated density jumps and the location of the CMB radius simultaneously. All the contours correspond to the sensitivity at $1\sigma$ (2 d.o.f) for 1~Mt$\cdot$yr exposure, assuming the true neutrino mass ordering to be NO. The top left (right) panel depicts the sensitivity contours for simultaneously constraining the density jump at the IC-OC (CMB) boundary and the location of the CMB radius. Similarly, the bottom left (right) panel shows the sensitivity for the density jump at the IM-MM (MM-OM) boundary, along with the location of the CMB radius. In each panel, the red (blue) contours show the expected sensitivity with (without) the CID capability of the ICAL detector. From Fig.~\ref{fig:2D_results}, we can observe that the neutrino oscillation data would be able to further constrain the allowed parameter space.

\section{Conclusions}

In this study, we exploit the weak interactions of neutrinos to probe the interior of the Earth, which provide independent and complementary information to what we get from traditional measurements such as seismic and gravitational studies. We demonstrate that using the Earth's matter effects in neutrino oscillations, atmospheric neutrino experiments like the INO-ICAL would be able to determine the density jumps at various boundaries and the location of CMB radius simultaneously.

\begin{acknowledgments}
We acknowledge financial support from the DAE, DST, DST-SERB, Govt. of India, INSA, and the USIEF. A.K.U. acknowledges financial support from the DST, Govt. of India (DST/INSPIRE Fellowship/2019/IF190755). The numerical simulations were performed using the SAMKHYA: High- Performance Computing Facility at the Institute of Physics, Bhubaneswar.
\end{acknowledgments}

\bibliographystyle{apsrev4-1}
\bibliography{References.bib}

\end{document}